# Slow Dephasing of Coherent Optical Phonons in Two-dimensional Lead Organic Chalcogenides

Hanjun Yang[1,2*], Sagarmoy Mandal[1], Bowen Li[3], Tushar Kanti Ghosh[1], Jonas Mark Peterson[1], Peijun Guo[3], Letian Dou[1,2], Ming Chen[1*], Libai Huang[1*]

[1]Department of Chemistry, Purdue University, West Lafayette, Indiana 47907, USA

[2]Davidson School of Chemical Engineering, Purdue University, West Lafayette, Indiana 47907, USA

[3]Department of Chemical and Environmental Engineering & Energy Sciences Institute, Yale University, New Haven, Connecticut 06516, USA

**Abstract**

Hybrid organic-inorganic semiconductors with strong electron-phonon interactions provide a programmable platform for developing a variety of electronic, optoelectronic, and quantum materials by controlling these interactions. However, in current hybrid semiconductors, such as halide perovskites, anharmonic vibrations with rapid dephasing hinder the ability to coherently manipulate phonons. Here, we report the observation of long-lived coherent phonons in lead organic chalcogenides (LOCs), a new family of hybrid two-dimensional semiconductors. These materials feature harmonic phonon dynamics despite distorted lattices, combining long phonon dephasing times with tunable semiconducting properties. Dephasing time as long as 75 ps at 10 K, with up to 500 cycles of phonon oscillation between scattering events, was observed, corresponding to a dimensionless harmonicity parameter more than an order of magnitude larger than that of halide perovskites. The phonon dephasing time is significantly influenced by anharmonicity and centrosymmetry, both of which can be tuned through the design of the organic ligands thanks to the direct bonding between the organic and inorganic motifs. This research opens new opportunities for the manipulation of electronic properties with coherent phonons in hybrid semiconductors.

**Introduction**

Hybrid organic-inorganic semiconductors, such as halide perovskites, represent an emerging class of highly tunable optoelectronic materials.[1-4] Electron-phonon coupling, including polaron formation, plays a crucial role in determining charge carrier transport and recombination in these materials.[5-7] Significant advances have been made in understanding how both the inorganic and organic components can modify electronic-phonon interactions. Beyond the thermally excited phonons with random phase and broad energy distribution, the utilization of coherent phonon with well-defined phase and energy is an intriguing approach to manipulate the material properties. For example, the wave-like interference of coherent phonons can be engineered in quantum wells and phononic crystals to manipulate the thermal conductivity,[8, 9] and coherent phonons can be conceived for producing terahertz (THz) frequency combs for quantum phononic engineering.[10] Strong coupling between exciton and acoustic phonon form the large polarons that transport coherently in van der Waals superatomic superlattices $Re_6Se_8Cl_2$ even at room temperature.[11, 12] A fundamental prerequisite for the manipulation of material properties through coherent optical phonons is a sufficiently low phonon dephasing rate, which determines the achievable coherence time.[13] However, due to the substantial anharmonicity of lattice vibrations and the dynamic lattice fluctuations, selective generation of optical phonon with slow dephasing rates in hybrid materials remains a challenge; for example, the coherent time is limited to a few ps in halide perovskites and silver phenylselenolate (AgSePh). [14-16]

Here we demonstrate 2D lead organic chalcogenide (LOCs), featuring van der Waals type layered structures with the inorganic Pb-S sublattice sandwiched between the organic thiol ligands (**Fig. 1a, b**),[17] as a new family of hybrid organic-inorganic semiconductors featuring coherent phonons with slow dephasing rate. The thiolate ligands permit the tuning of the dimensionality and inter-plane interactions, thereby modifying the phonon energy dispersion and dephasing. Importantly, these structures differ from the extensively studied lead halide 2D perovskites by the direct Pb-S bonds between the inorganic and organic sublattices, providing new opportunities to design vibrational modes and symmetry using organic ligands. LOCs belong to metal organic chalcogenides (MOCs), a broadly defined family of hybrid semiconductor materials composed of metal cations and coordinating organic thiol, selenol or tellurol.[18-20] Thus far, AgSePh and its derivatives are the most studied MOCs with anisotropic excitons and strong light-matter interaction.[21-24] Compared to AgSePh , the stereochemically expressed $6s^2$ lone pair of $Pb^{2+}$ in

LOCs result in highly distorted coordination structures and hence strong electron-phonon coupling as observed in several Pb-based materials.[20, 25, 26] Therefore, these LOCs present a unique opportunity to systematically investigate the interactions between the electronic and lattice degrees of freedom in soft 2D hybrid materials to optimize phonon coherence.

We measured the dephasing dynamics of several 2D LOCs by temperature-dependent coherent phonon spectroscopy, where ultrafast laser excitations generate coherent phonons by transferring the phase of laser pulse to lattice vibrations. Unlike thermally excited phonons with random phases, these coherent phonons exhibit emergent photophysical properties. Coherent phonon spectroscopy has been performed in a wide range of materials, such as metals and semi-metals,[27] transition metal chalcogenides,[28-30] halide perovskites,[14, 26, 31, 32] MOCs,[15, 16] and semiconductor heterostructures.[33] The dephasing of coherent phonons is typically associated with the phonon down-conversion mediated by lattice anharmonicity, namely the vibrational energy deviation from quadratic relationship with regard to lattice displacement.[34] Hybrid semiconductors often feature soft, flexible bonds and short-range polaronic deformation potential, resulting in dynamic distortion and large anharmonicity.[15] The anharmonicity is especially strong in halide perovskites due to the weak Pb-halide bonds, and the stereochemically expressed Pb $6s^2$ lone pairs with flat potential surface.[35] In comparison, LOCs feature a more rigid Pb-S network with static distortion that should form a more harmonic phonon energy surface despite the seemingly distorted crystal structures.

Our experiments demonstrate that the hemidirected LOCs exhibit low-energy coherent optical phonons with dephasing times exceeding 75 ps. The slow dephasing time of LOCs results from the synergistic effects of low phonon energy, low anharmonicity, and suppressed optical phonon scattering. These findings are supported by ultrafast spectroscopy and low-frequency Raman scattering, and further corroborated by first-principles calculations, which reveal highly isolated, low-energy optical phonons. Our results established 2D LOCs as a configurable platform for the study of electron-phonon coupling and other wave-like phonon properties in 2D hybrid semiconductors.

**Results**

**Coherent optical phonon generation in LOCs:** Three quasi-2D and 2D LOCs were mainly studied in this report, with the increasing Pb-S bond connectivity and an associated decrease in the band gap from lead 4-(methoxycarbonyl)benzenethiolate (Pb(MMBA)$_2$), to lead 4-(methylthio)benzenethiolate (Pb(MSBT)$_2$), and lead 4-methoxybenzenethiolate (Pb(MOBT)$_2$) (**Fig. 1a, b**, Fig. S1). Their synthesis and basic optical properties have been reported in prior reports.[17, 36] Transient reflectance (TR) spectroscopy was used to characterize the ultrafast carrier and phonon dynamics. Compared with steady-state measurements such as infrared and Raman spectroscopy, the pump-probe experiments track the phonon population in the time domain, resolving the phonon dephasing processes.[7, 13, 37] In addition, the pump-probe experiments selectively characterize phonons that couple to electronic transitions, serving as a direct measurement on the electron-phonon coupling effects. Herein we present Pb(MSBT)$_2$ as an example and **Fig. 1c** shows a representative 2D pseudo-color map of the room temperature TR transient spectra. The TR map features a strong derivative peak around 550 nm and a weaker, broad peak around 670 nm. (Fig. S2). These two signals are tentatively assigned to the photoinduced refractive index change near the direct and indirect band gap transitions, respectively. Both peaks exhibit fast recovery rate on the order of ~30 ps (Fig. S2) at room temperature, likely a result of rapid self-trapping which has been observed in the AgTePh.[38] The self-trapped species exhibit low absorption and thus are hardly detectable in the TR spectra, despite exhibiting ns to μs lifetime and strong luminescence at low temperature.[39] The other two LOC compounds, Pb(MMBA)$_2$ and Pb(MOBT)$_2$, feature qualitatively similar TR spectra composed of two derivative peaks, despite of their different degrees of distortion and therefore different band gap energies (Fig. S3, 4). The similar carrier dynamics in three LOCs originate from their overall similar electronic bands structures.[17] Notably, the TR peaks overall slightly shift to higher energy with increasing delay time, most obvious for Pb(MOBT)$_2$ (e.g. from 576 nm at 1 ps to 564 nm at 800 ps, Fig. S4), showing extensive band gap renormalization at high carrier concentration.

A closer look at TR map reveals a beating pattern, the signature of coherent phonons (**Fig. 1c**, inset). The in-phase phonon oscillation (whose phase is defined by the ultrafast laser) modulates the complex refractive index of LOC samples, resulting in a periodic pattern in the TR kinetics (Fig. S5).[40] The coherent phonon spectra can be obtained by subtracting the decaying component from the TR spectra at each wavelength, revealing the intrinsic phonon frequency (**Fig. 1d**). The Fourier transformation (FT) of the obtained TR residue (**Fig. 1e**) shows that only one phonon mode is strongly coupled to the electronic transition in Pb(MSBT)$_2$ an Pb(MOBT)$_2$, while a weak high-frequency peak can be observed as well in the Pb(MMBA)$_2$ sample, consistent with its Raman spectrum (*vide infra*). The beating patterns of the TR residue of all LOC compounds were fit by simple exponential damped cosine function $I = I_0 \cos\frac{(t-t_0)}{t_{os}} \exp\left(-\frac{t}{t_d}\right)$ to extract the phonon dephasing kinetics. While all three LOCs showed clear evidence of the coherent phonons, the beating period $t_{os}$ as well as the damping time $t_d$ exhibit strong dependence on the crystal structure of the materials (**Fig. 1f**). Specifically, the phonon beating period is inversely proportional to the level of lattice distortion (Table S1), from 1.416 ± 0.006 ps (23.5 ± 0.1 cm$^{-1}$) for Pb(MMBA)$_2$ and 1.803 ± 0.003 ps (18.47 ± 0.03 cm$^{-1}$) for Pb(MSBT)$_2$, to 5.67 ± 0.03 ps (5.59 ± 0.03 cm$^{-1}$) for Pb(MOBT)$_2$. Additionally, the dephasing times increase from 7.4 ± 1.0 ps for Pb(MMBA)$_2$ to 12.2 ± 0.9 ps for Pb(MSBT)$_2$ and 15.5 ± 1.2 ps for Pb(MOBT)$_2$. Compared with layered hybrid materials such as 2D halide perovskites and AgSePh MOCs, the highly distorted LOCs exhibited surprisingly long dephasing times even at room temperature (Table S2)[15, 16, 28, 30, 31, 41-44].

The low phonon frequencies raise questions regarding whether the observed coherent phonons belong to the optical or acoustic branch. The deformation potential mechanism, a common mechanism for coherent acoustic phonon (CAP) in semiconductors, can be summarized by the refractive index modulation due to a propagating strain wave caused by the optical pump, which then experience optical interference.[45-47] This mechanism predict a strong dependence of the beating frequency on the probe wavelength.[48] Noting that the oscillating frequencies of all LOCs samples are independent of the probe wavelength (Fig. S2, S3, S4), we exclude the relevance of CAP caused by thermoelastic stress and deformation potential. On the other hand, the coherent phonon frequencies of LOCs resemble the low energy phonon mode in the corresponding high-resolution Raman spectra (**Fig. 2a**). Therefore, we conclude that the low-energy optical phonon is the dominant species in the electron-phonon coupling in LOCs. A closer look at the Raman spectra reveals that both Pb(MSBT)$_2$ and Pb(MOBT)$_2$ exhibit an isolated single optical phonon mode at low-frequency range, suggesting limited phonon-phonon scattering. Remarkably, sharp Raman peaks were observed even at room temperature (**Fig. 2a**), in contrast to halide perovskites where Raman peaks broaden and diminish near room temperature due to the dynamically disordered lattice.[49] These results suggest that the lattices in the LOCs are more rigid and features more harmonic vibrations compared to lead halide perovskites.

Coherent optical phonons can be generated by two representative mechanisms, the impulsive stimulated Raman scattering (ISRS) and displacive excitation of coherent phonon

(DECP) [27, 50]. ISRS can be described as a field-driven phonon generation, where the phonons are generated by Raman scattering stimulated by pump pulses that are much faster than phonon oscillation time. The equilibrium atomic coordination remains unchanged in the ISRS mechanism and the excited lattice forms a sine-type oscillation pattern. On the other hand, the DECP mechanism, which is most prominent in Peierls-distorted metals such as Bi and Te, describes a nuclear displacement of the equilibrium coordinate under optical excitation.[27, 51] This rapid displacement caused by photocarrier generation initiates the coherent phonons, which in turn modulates the dielectric function and affects the observed transient reflectance signal. We attribute the observed coherent phonons in LOCs to the DECP mechanism.[27] The cosine-type oscillation, a signature of DECP mechanism, is clearly illustrated near time zero (Fig. S6). Further, the pump-probe experiments with below band gap excitation (785 nm) with even higher pump fluence yield little to no TR signal despite increased pump intensity (Fig. S7), showing the necessity of electronic excitation in the coherent phonon generation.

DFT calculations were employed to furtherly elucidate the nature of these coherent phonon modes. All LOCs investigated exhibit a pair of low energy phonon modes with A and B symmetry ($A_g$ and $B_g$ modes for Pb(MOBT)$_2$, **Fig. 2b** and Fig. S8-S10) with similar atomic displacement pattern. The A ($A_g$) modes correspond to the scissoring modes along the *b* axis, which align parallelly with the long Pb-S bonds, whereas the B ($B_g$) modes involve scissoring motion along the *a* axis, which is roughly orthogonal to the long Pb-S bonds (**Fig. 2c**). Overall, the calculated phonon energies match the experimental values obtained by coherent phonon measurement as well as Raman spectroscopy (**Fig. 2b**). The band gap shifts $\Delta E$ as functions of the atomic displacement were calculated to assess the electron-phonon coupling in LOCs (**Fig. 2d,** Fig. S11). The A modes lead to an asymmetric change in $\Delta E$ near the origin, implying that the A phonons induce significant band modulation at the phonon frequency. On the other hand, the B modes induce symmetric $\Delta E$ with respect to positive and negative displacement, and $\Delta E$ that are 1~2 orders of magnitude smaller coupling, as a result of the degeneracy linked by symmetry operations.[14] In summary, the DFT calculations show that the electronic transitions in LOCs are strongly coupled with the generation of fully symmetric A ($A_g$) phonons modes.

**Optical phonon dephasing dynamics**: Compared with steady-state vibrational spectroscopy, probing the coherent phonons with pump-probe spectroscopy provided insights into the relaxation and dephasing dynamics of the phonons as well as electron-phonon coupling. We investigated the temperature-dependent behavior to study the effect of population distribution of optical phonons. **Fig. 3a-c** summarizes the coherent phonon kinetics of the three LOCs at low temperatures. All samples showed increased phonon energy at reduced temperature, agreeing well with the low-temperature Raman spectra (**Fig. 3d,** Fig. S12). The agreement between non-resonant Raman and TR results suggests that phonon energy shift is mainly the result of the lattice contraction due to third-order anharmonic phonon softening. In conjunction with the phonon frequency, the dephasing time significantly increased at low temperatures due to the suppression of phonon-phonon scattering. Most strikingly, Pb(MOBT)$_2$ exhibits > 75 ps phonon coherence time at 10 K, which is one of the longest-lived coherent optical phonons among hybrid 2D materials (**Fig. 3e**, Table S2).

We apply a cubic overtone model to fit the temperature-dependent phonon energy and dephasing rate as shown in **Fig. 3d-f**, where an optical phonon down-convert to two acoustic phonons satisfying both energy and momentum conservations.[52, 53]

$$\omega = \omega_0 + \gamma_\omega \left(1 + \frac{2}{\exp(\hbar\omega_0/2k_BT) - 1}\right)$$

$$\Gamma = \Gamma_0 + \gamma_d \left(1 + \frac{2}{\exp(\hbar\omega_0/2k_BT) - 1}\right)$$

Where $\omega$ and $\omega_0$ represent the measured and zero-temperature optical phonon wavenumber, $\Gamma$ and $\Gamma_0$ represent the measured dephasing rate and the zero-temperature dephasing rate, $\gamma_\omega$ and $\gamma_d$ are constants representing the anharmonic effect on phonon wavenumber and dephasing rate. Pb(MSBT)$_2$ data below 120 K were not used in the analysis (*vide infra*). This model fits the data reasonably well, suggesting that three-phonon processes are the primary mechanism for anharmonic relaxation. The fitted phonon wavenumber anharmonic parameter $\gamma_\omega$ are $(1.12 \pm 0.05) \times 10^{-1}$ cm$^{-1}$, $(4.6 \pm 0.9) \times 10^{-2}$ cm$^{-1}$, and $(1.4 \pm 0.1) \times 10^{-2}$ cm$^{-1}$ for Pb(MMBA)$_2$, Pb(MSBT)$_2$ and Pb(MOBT)$_2$, respectively. The dephasing rate anharmonic parameters $\gamma_d$ are $(2.9 \pm 0.7) \times 10^{-3}$ ps$^{-1}$, $(1.3 \pm 0.4) \times 10^{-3}$ ps$^{-1}$, and $(5.1 \pm 0.5) \times 10^{-4}$ ps$^{-1}$ for Pb(MMBA)$_2$, Pb(MSBT)$_2$ and Pb(MOBT)$_2$, respectively. A unitless parameter $\omega_0/(2\pi\gamma_d)$ that is interpreted as the average cycles of phonon oscillation between scattering events is used to compare the harmonicity of phonon modes with varying energy,[53] being $2.8\times10^2$, $4.7\times10^2$, and $4.7\times10^2$ respectively for Pb(MMBA)$_2$, Pb(MSBT)$_2$ and Pb(MOBT)$_2$, respectively. Comparing with hybrid lead halide perovskites and AgSePh, the LOCs exhibit an order of magnitude lower anharmonicity.[15, 53] In addition, short time Fourier transform (STFT) revealed no systematic shift for Pb(MSBT)$_2$ and Pb(MOBT)$_2$ samples at all temperatures, while Pb(MMBA)$_2$ can exhibit a slight shift with time to high frequency at high temperature, consistent with its larger degree anharmonicity (Fig. S13-S16). In summary, despite the asymmetric Pb-S bonds, the anharmonic effect on phonon energy is significantly suppressed, resulting in the long-lived coherent phonons.

Unlike the other two compounds where the Fourier transform of coherent phonon largely remains unchanged at low temperatures, a new phonon mode at high wavenumber of 53 cm$^{-1}$ emerged in Pb(MSBT)$_2$ at below 120 K (Fig. S13, S17). The dephasing time of this high energy phonon was estimated to be 4.5 ± 0.7 ps at 10 K, 7.0 ± 1.2 ps at 30 K, and 7.4 ± 1.3 ps at 60 K based on single exponential fitting of the STFT amplitude (Fig. S17). At higher temperatures the STFT amplitudes are too weak to be resolved. A phonon mode around 53 cm$^{-1}$ is observed in Raman spectra at all temperatures as well, albeit being rather weak (Fig. S17). DFT calculations attributed this phonon mode to a symmetric stretching of Pb-S octahedral coupled with in-plane ligand translation, with an energy of 52.91 cm$^{-1}$ (Fig. S9). The emergence of high energy coherent phonons modes at <120 K further scatter with the lowest energy coherent modes, thus invalidating the three-phonon scattering model where scattering with acoustic phonon was assumed to be the main dephasing mechanism.

**The role of structure and symmetry in phonon dephasing**: Next, we investigate how the structure and symmetry affect phonon dephasing dynamics. Qualitatively, the calculated vibrational energy (Fig. S18) of the coherent phonon modes of all samples show close to parabolic curves, revealing their high degree of harmonic oscillation. Overall, the long dephasing time of Pb(MOBT)$_2$ cannot be solely attributed to low anharmonicity. The crystal symmetry affects the vibrational selection rule, which may affect the observed phonon dephasing. Unlike the other LOCs, Pb(MOBT)$_2$ has a centrosymmetric unit cell with P2/c space group. To decouple the structural and symmetric factors within LOCs, we compared Pb(MOBT)$_2$ to Pb(EOBT)$_2$, an additional LOC with similar local Pb-S coordination (Fig. S19, crystal structure in CCDC# 2340737). Featuring an ethoxy instead of the methoxy group, the EOBT ligand exhibits crystal structures and chemical properties similar to Pb(MOBT)$_2$ except that Pb(EOBT)$_2$ forms a non-centrosymmetric unit cell (P2$_1$). Surprisingly, the Pb(EOBT)$_2$ exhibits drastically weaker phonon modulation and shorter-lived coherent phonons. The coherent phonons can only be observed below 90 K, at significantly higher phonon energy at 15.5 cm$^{-1}$ and 28.9 cm$^{-1}$, and with much faster phonon dephasing time at < 8 ps. The phonon frequencies are consistent with those measured by Raman spectroscopy (**Fig. 4a, b**, Fig. S20, S21). DFT calculation shows a pair of A phonons at 25.5 cm$^{-1}$ and 32.1 cm$^{-1}$, corresponding to the *b*-axis scissoring and out-of-plane inorganic sublattice displacements (Fig. S22). In addition, the Raman peaks are qualitatively broader than all other LOCs, reflecting a higher level of anharmonicity. DFT calculations suggest that the vibrational potential anharmonicity for Pb(MOBT)$_2$ and Pb(EOBT)$_2$ (**Fig. 4c**) is similar as well. Given the otherwise similar properties between Pb(MOBT)$_2$ and Pb(EOBT)$_2$, we concluded that the centrosymmetry of Pb(MOBT)$_2$ could lead to selection rules that limit phonon scattering, playing an important role in suppressing the dephasing of coherent phonon.

Further, unlike halide perovskites where ligands are coupled indirectly to the inorganic sublattice, Pb-S bonds between the organic and inorganic components in LOCs can lead to ligand movements and the spatial hindrance that strongly influence the phonon transitions. We note that the methoxy group in Pb(MOBT)$_2$, unlike other LOCs, slightly interdigitate into adjacent layers thus resulting in greater restrictions of vibration. Careful examination of the atomic displacement contribution confirms a significantly reduced ligand side-chain contribution in Pb(MOBT)$_2$, thus hypothetically reducing the dissipation via ligand side chains (**Fig. 4d**) and contribute also to the suppressed phonon-phonon scattering. Thus, we rationalized that the longer-lived coherent phonons in LOCs especially Pb(MOBT)$_2$ is a combined effect of low anharmonicity and low phonon energy, centrosymmetry, and strong interlayer ligand interaction.

Electron-phonon coupling is also manifested as phonon energy modulation resulting from increase in carrier density, a phenomenon known as electronic softening. The charge-transfer type electronic excitation in LOCs weakens the Pb-S bond, resulting in decreased phonon energy (Fig. S23). Using Pb(MOBT)$_2$ at 10 K as an example, the A$_g$ Phonon energy experienced a 2.3% decrease from 7.97 cm$^{-1}$ to 7.79 cm$^{-1}$ as carrier density increases. The phonon energy as a function of delay time does not significantly shift over time as revealed in STFT spectra, confirming a static softening on contrary to a dynamic anharmonic effect (Fig. S24).[54] However, a drastic decrease in the phonon dephasing time is observed from ~120 ps to 33 ps with increasing carrier density because of the increased phonon-phonon scattering rate, originating from the unavoidable increasing phonon population. It should be expected that further decreasing the coherent phonon density may even further increase the dephasing time > 75 ps, but the accurate measurement is limited by the current measurement sensitivity. We note that the electron-phonon scattering and phonon-phonon scattering coexist in the pump-probe experiments, and additional experiments may be needed to quantify each contribution.[55]

**Discussion**

LOCs with direct chemical bonds between organic and inorganic components provide new opportunities for utilizing long-lived coherent phonons in hybrid semiconductors. The electronic properties of these materials are highly susceptible to the manipulation through small displacements of the Pb-S bonds between the organic and inorganic components, which can be controlled optically using ultrashort light pulses. Here, we compare the phonon dephasing in LOCs to other 2D semiconductors, prerequisite for realization of such manipulation. The dimensionless coherent parameter $\omega_0/(2\pi\gamma_d)$ is ~ 500 for LOCs, indicating that the optical phonons oscillate for about 500 cycles before dephasing. This value is more than one order of magnitude larger than that of 2D halide perovskites,[53, 56] where carriers and excitons typically reside in the inorganic motifs. 2D halide perovskites generally suffer from a large degree of dynamic distortion and dense low-energy vibrational modes, leading to rapid phonon dephasing. LOCs improve upon halide perovskites by constructing the crystal framework with direct bonds between the organic and inorganic motifs. This rigid framework reduces phonon anharmonicity and enhances the tunability of phonons through ligand design. The phonon coherence in LOCs is similar to those found in transition metal dichalcogenides (TMDCs), where coherent phonons have been proposed as a means to control electronic, optical, and magnetic properties.[57] However, compared to TMDCs, LOCs offer additional advantages, including structural tunability and the ability to design symmetry and dimensionality with organic ligands. These features make LOCs a promising platform for designing hybrid semiconductors with tunable coherent phonon properties.

In summary, we reported a joint experimental and theoretical investigation of coherent phonons in LOCs. The strong electron-phonon coupling in LOCs facilitates the generation of coherent optical phonons via the DECP mechanism. Organic ligands have a profound influence on the energy and dephasing kinetics, yielding long-lived coherent phonons with dephasing times exceeding 75 ps at 10 K. Interestingly, the observation that highly distorted LOC materials exhibit highly harmonic phonon modes calls for a reevaluation of the role of lattice distortion in phonon spectra. These findings point to new opportunities to explore soft materials with static deformation in search of unique coherent phonons properties, enabling optical manipulation of materials properties in a coherent and designable manner. For example, the coherent vibrations in LOCs govern polaronic lattice deformation, potentially reducing hopping energy barriers and accelerating polaron transport.[58]

## Methods

**Sample preparation.** MMBA was obtained from Combi-Block. EOBT was obtained from Enamine. All other chemicals were obtained from Sigma-Aldrich. All chemicals were used as received. The syntheses of Pb(MMBA)$_2$, Pb(MSBT)$_2$, and Pb(MOBT)$_2$ have been reported previously.[17] Pb(EOBT)$_2$ powder was obtained similarly by the reaction between 0.6 mmol lead acetate trihydrate and 1.25 mmol EOBT. Pb(EOBT)$_2$ single crystals were obtained by dissolving 3 mg Pb(EOBT)$_2$ powder in a mixture of 150 μL of γ-butyrolactone, 20 μL of dimethyl sulfoxide and 50 μL of ethanol at 100 °C and slowly cooled to room temperature in 3 days. All samples were exfoliated by scotch tape and transferred to Si/SiO$_2$ (500 nm) substrates for optical measurements. Crystallography data of Pb(EOBT)$_2$ was deposited in Cambridge Crystallographic Data Centre (CCDC) with deposition number 2340737.

**Ultrafast spectroscopy measurements.** Transient reflection spectrum was measured using a femtosecond pump-probe system based on a home-built confocal microscope. 1030 nm laser pulses with 250 fs pulse duration was generated by a 250 kHz amplified Yb:KGW laser (Pharos, Light Conversion). Continuum probe beam was generated by focusing 5% of the fundamental 1030 nm laser onto a YAG crystals. The rest of the fundamental beam was used to pump an optical parametric amplifier (OPA, TOPAS-Twins, Light Conversion) as the pump beam at 470 nm, which is further modulated by a chopper at 360 Hz. The pump and probe beam were then combined and focused onto the sample using a 40× objective (NA = 0.6). The reflected beam was collected by the same objective and analyzed by a spectrometer module (Exemplar LS, BWTek). Low-temperature measurements were conducted on a Montana s50 cyrostation. A triexponential decay was fitted and subtracted at each wavelength for the resulting TR spectrum to obtain the coherent phonon oscillation signal.

**Raman spectroscopy.** Raman measurements were conducted using a custom-build microscope-integrated spectrometer system.[59] The specimen was place in an optical cryostat (Janis VPF-100), which was then vacuum pumped to maintain pressure below $1\times10^{-4}$ Torr. A temperature controller (LakeShore 330) was connected to the cryostat while liquid nitrogen was added to enable cooling down to 77 K. A long-working range objective lens (Mitutoyo, NIR, 20x, NA=0.4) was utilized to focus a frequency-stabilized 785 nm excitation laser (Toptica) onto the sample. The reflected Raman signal was collected by the same objective lens and was filtered by a set of five narrow-linewidth, reflective volume Bragg grating notch filters (OptiGrate), which allowed measurements of Raman signals as low as 5 cm$^{-1}$. The Raman signal then passed through a pair of 75-mm focal length achromatic lens and a 50 μm pinhole, entered the spectrograph (Horiba iHR550, grating 950 Grooves/nm), and was resolved by a CCD camera (Syncerity UV-Vis).

**Calculations.** All the calculations were performed within the framework of periodic density functional theory (DFT) employing the Quantum ESPRESSO package.[60] Perdew–Burke–Ernzerhof (PBE)[61] functional at the level of generalized gradient approximation (GGA) was used together with projector augmented wave (PAW)[62] type pseudo-potentials. Grimme's D2 correction[63] was incorporated to account for dispersion interactions. The wavefunctions and electron densities were expanded in a plane wave (PW) basis set with cutoff of 60 Ry and 300 Ry, respectively. Convergence of the wavefunctions was achieved when the magnitude of change in energy fell below $10^{-10}$ Ry. The convergence criteria for the optimization of lattice parameters and the geometries were set to 0.5 kbar and $10^{-4}$ Ry/au, respectively. The Brillouin zone was sampled with Γ-centered k-point mesh of 4×4×1 for Pb(MOBT)$_2$ and Pb(EOBT)$_2$ systems. For Pb(MMBA)$_2$ and Pb(MSBT)$_2$ systems, Γ-centered k-point mesh of 3×3×1 was used. Starting from the relaxed structures, phonon calculations were performed for all compounds using density functional perturbation theory (DFTP).[64] Finally, acoustic sum rule (ASR) was applied to remove small negative frequencies from the normal modes at Γ point. To understand the relation between band gap shift and phonon displacement, the band gap was calculated as a function of positive and negative atomic displacements along specific phonon modes. Similarly, the potential energy surfaces were calculated along different phonon modes following the same methodology.

## Acknowledgements

The work at Purdue is supported by the US Department of Energy, Office of Basic Energy Sciences under award number DE-SC0022082. Ultrafast spectroscopy instrumentation was partly supported by National Science Foundation through award NSF-CHE-2117616. Authors thank Matthias Zeller from Department of Chemistry, Purdue University for the single crystal crystallography. Work at Yale University is supported by the National Science Foundation under Grant No. DMR-2313648. M. C., T.K.G. and S. M. would like to acknowledge financial support from the NSF CSEDI EAR-2246687. Computational resources were provided by Anvil at Purdue University through allocation CHE220008 from the Advanced Cyberinfrastructure Coordination Ecosystem: Services & Support (ACCESS) program, which is supported by National Science Foundation grants #2138259, #2138286, #2138307, #2137603, and #2138296. H.Y. acknowledges financial support from the Lillian Gilbreth Postdoc Fellowship by the college of engineering of Purdue University.

**Author contributions**

H. Y., L. D. and L. H. conceived the project. H. Y. conducted sample synthesis, transient reflection measurements and conducted data analysis. B. L. and P. G. carried out Raman spectroscopy and analysis. S. M., T. K. G. and M. C. carried out DFT calculation and analysis. The manuscript is written through contributions from all authors. All authors have given approval to the final version of the manuscript.

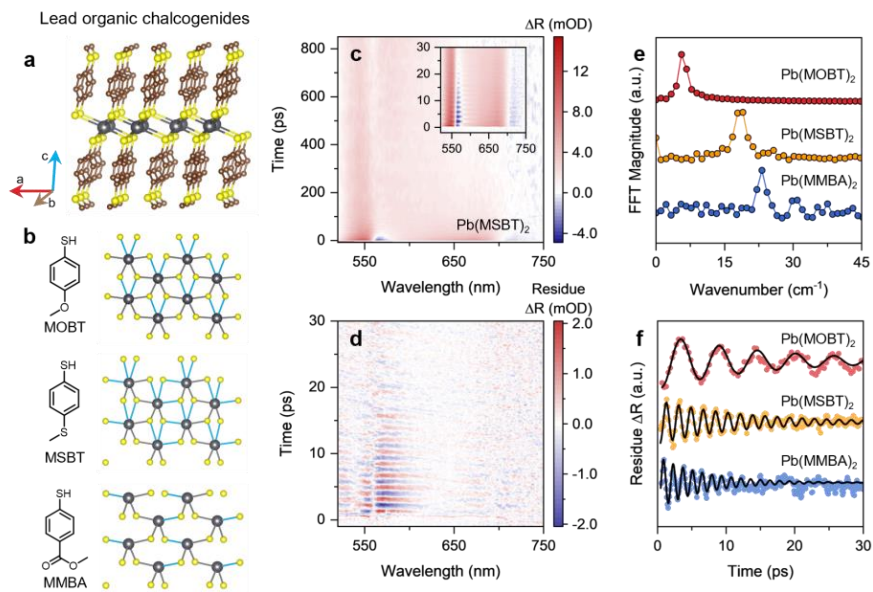

**Fig 1. a**, Schematic illustration of the LOC crystal structure using Pb(MSBT)$_2$ as an example. **b**, Ligand structures and the corresponding LOC in-plane crystal structures. Blue bonds represent long Pb-S bonds (> 3.1 Å) and grey bonds represent short Pb-S bonds (< 3.1 Å). **c**, TR pseudo-color map of Pb(MSBT)$_2$ at room temperature. Inset: zoom-in view in the short delay time range. **d,** Oscillating component of the TR residue after subtracting exponential signal. **e**, FT spectra of the coherent phonon beating spectra. **f**, Normalized coherent phonon beating of different LOC compounds. Solid lines are fitted with single-exponential damped oscillator.

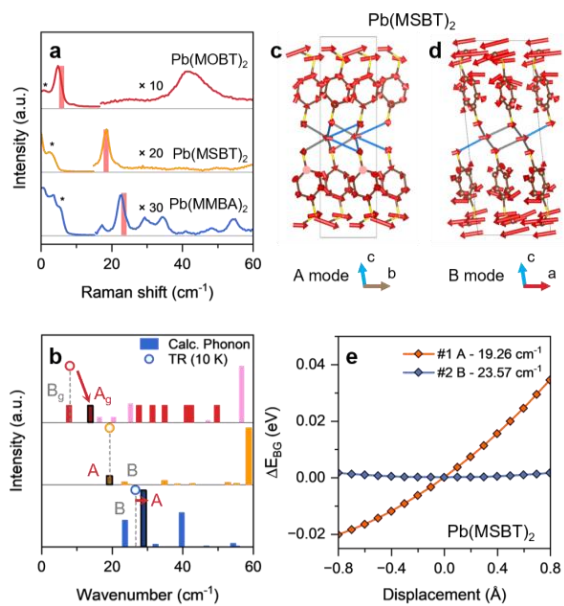

**Fig 2. a.** Low frequency Raman spectra of LOCs at room temperature. Coherent phonon wavenumber measured by TR are marked in shaded bars. Background signals are marked with asterisks. **b.** Calculated infrared intensity at Γ-point. Considering the centrosymmetry of Pb(MOBT)$_2$, the IR-active modes are marked in pink and the Raman-active modes (assigned with arbitrary intensity due to calculational limit) are marked in red. Coherent phonon wavenumber measured by TR spectroscopy at 10 K are marked in empty circles. The DFT calculated phonon modes corresponding to the TR spectra are highlighted. **c-d.** Schematic illustration of the A and B phonon modes of Pb(MSBT)$_2$. **e,** Calculated band gap change of Pb(MSBT)$_2$ as a function of displacement along the phonon eigenvectors.

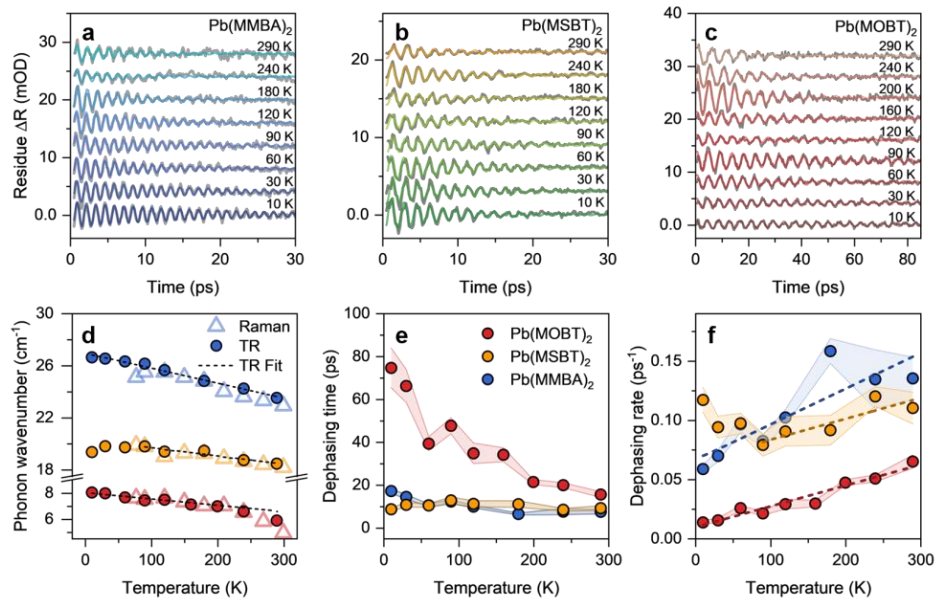

**Fig 3. a-c.** Temperature-dependent TR coherent phonon spectra of LOCs. **d,** Phonon wavenumbers as the function of temperature. Results based on low-frequency Raman spectroscopy are marked in triangles and results based on TR are marked in circles. Fitting curves are marked in dashed lines. **e, f.** Phonon dephasing time and phonon dephasing rate. Fitting curves are marked in dashed lines.

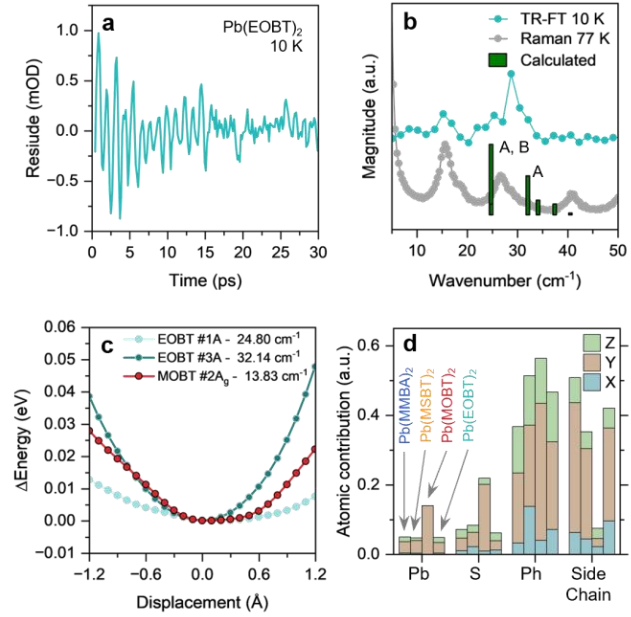

**Fig 4. a.** TR spectrum residue of Pb(EOBT)$_2$ at 10 K. **b.** FT of coherent phonon spectrum at 10 K, Raman spectrum at 77 K, and DFT calculated infrared intensity at Γ-point of Pb(EOBT)$_2$. **c.** Calculated vibrational energy potential of Pb(MOBT)$_2$ and Pb(EOBT)$_2$. **d**. Calculated atomic displacement amplitude for the corresponding coherent phonon modes. Amplitudes of Pb(EOBT)$_2$ are based on the optical phonon mode at 24.80 cm$^{-1}$.

**Slow Dephasing of Coherent Optical Phonons in Two-dimensional Lead Organic Chalcogenides**

Hanjun Yang[1,2*], Sagarmoy Mandal[1], Bowen Li[3], Tushar Kanti Ghosh[1], Jonas Mark Peterson[1], Peijun Guo[3], Letian Dou[1,2], Ming Chen[1*], Libai Huang[1*]

[1]Department of Chemistry, Purdue University, West Lafayette, Indiana 47907, USA

[2]Davidson School of Chemical Engineering, Purdue University, West Lafayette, Indiana 47907, USA

[3]Department of Chemical and Environmental Engineering & Energy Sciences Institute, Yale University, New Haven, Connecticut 06516, USA

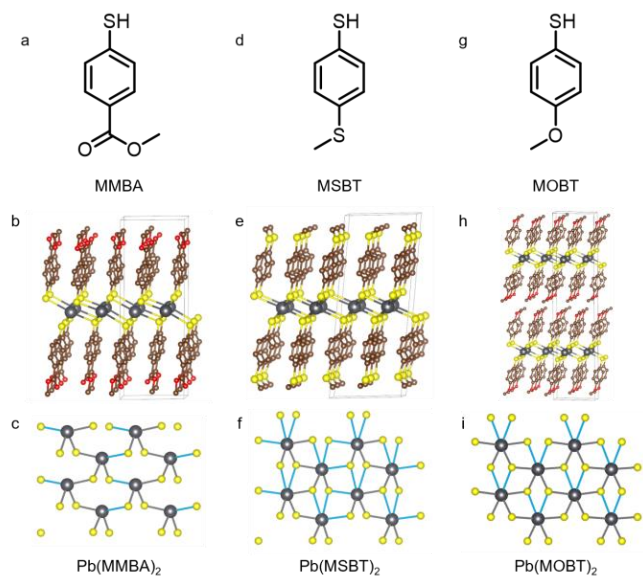

**Fig. S1. a, d, g**. Ligand structures. **b, e, h**. crystal structures of LOCs. **c, f, i**. In-plane structures of the inorganic sublattices of LOCs. Blue bonds in represent long Pb-S bonds between 3.1 Å to 3.6 Å. Grey bonds represent short Pb-S bonds < 3.1 Å.

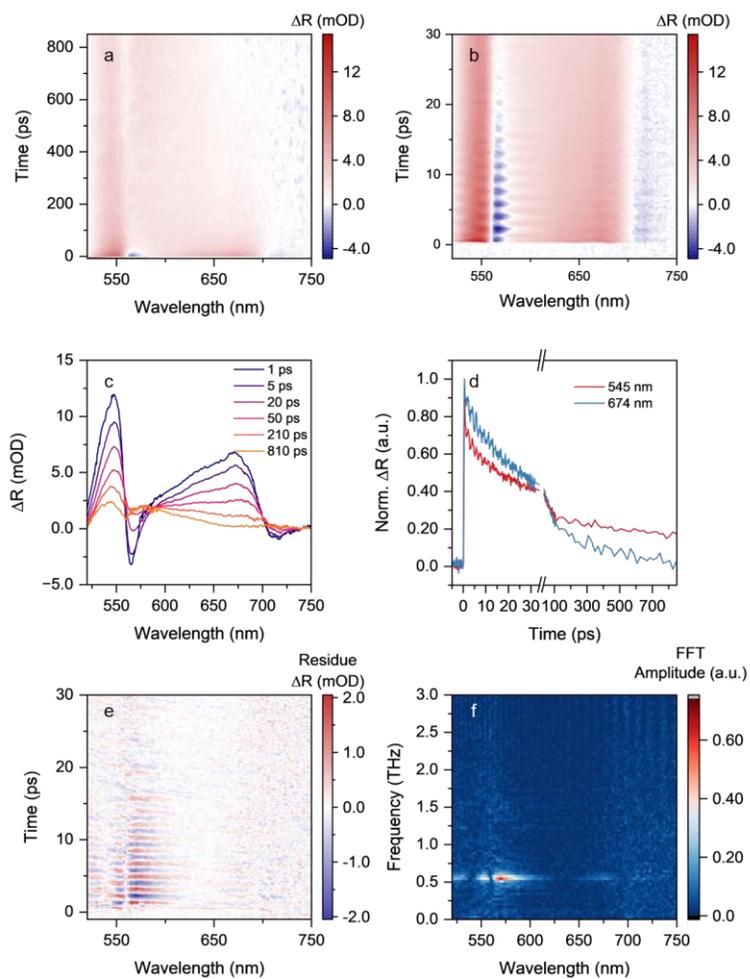

**Fig. S2. a, b.** TR pseudo-color map of Pb(MSBT)$_2$. **c.** TR spectrum of Pb(MSBT)$_2$ at different delay time. **d.** TR kinetic traces, probed at 545 nm and 674 nm. **e.** Pseudo-color map of residue TR signal. **f.** FT mapping of the residue TR signal from 0.5 to 30 ps.

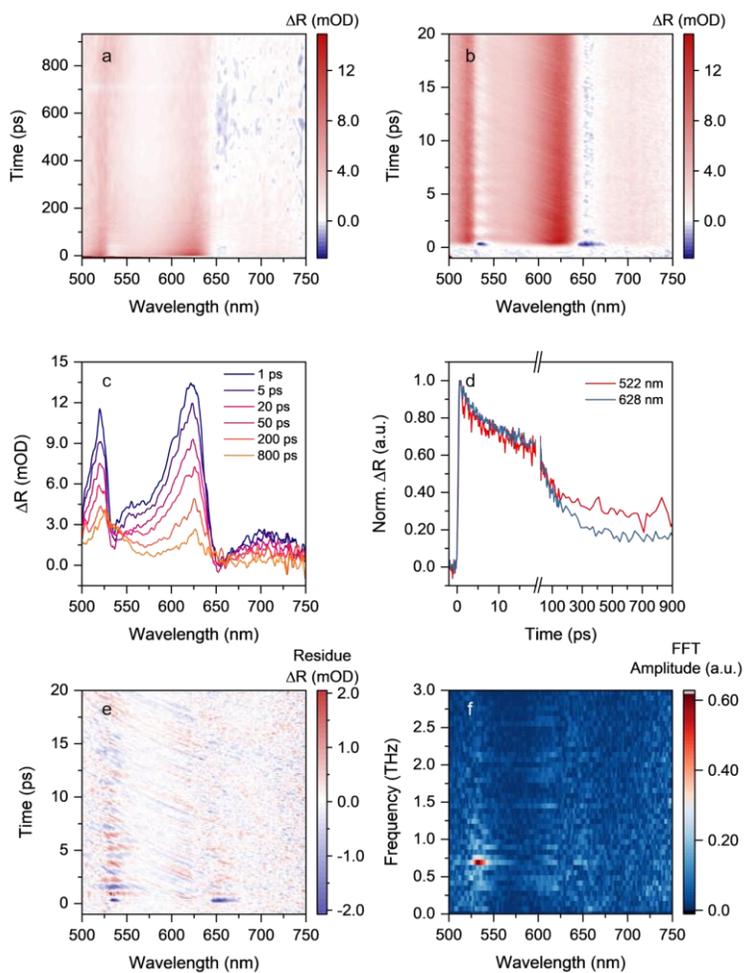

**Fig. S3. a, b.** TR pseudo-color map of Pb(MMBA)$_2$. **c.** TR spectrum of Pb(MMBA)$_2$ at different delay time. **d.** TR kinetic traces, probed at 522 nm and 628 nm. **e.** Pseudo-color map of residue TR signal. **f.** FT mapping of the residue TR signal from 0.5 to 15 ps.

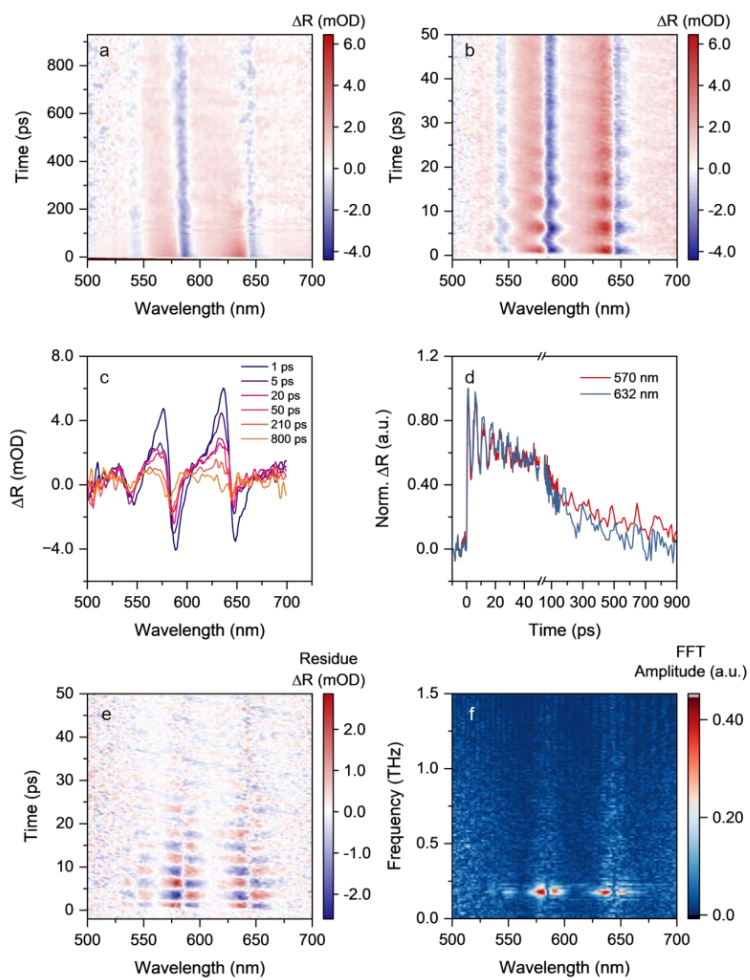

**Fig. S4**. **a, b**. TR pseudo-color map of Pb(MOBT)$_2$. **c.** TR spectrum of Pb(MOBT)$_2$ at different delay time. **d.** TR kinetic traces, probed at 570 nm and 632 nm. **e.** Pseudo-color map of residue TR signal. **f.** FFT mapping of the residue TR signal from 0.5 to 80 ps.

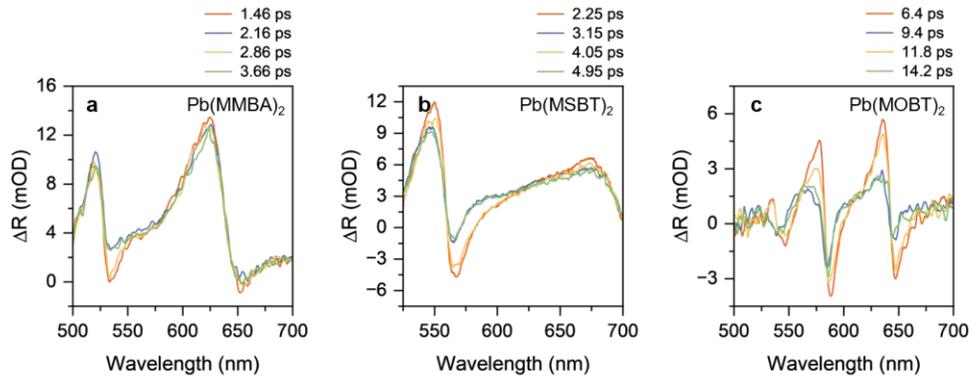

**Fig. S5**. Coherent phonon modulated TR spectra at room temperature.

**Table S1.** Distortion parameters of LOCs.

| Material | Pb(MMBA)$_2$ | Pb(MSBT)$_2$ | Pb(MOBT)$_2$ | Pb(EOBT)$_2$ |
|---|---|---|---|---|
| Δ | 0.017 | 0.010 | 0.0043 | 0.0065 |
| Σ (degree) | 17.10 | 15.49 | 15.49 | 15.38 |

Δ and Σ represent the bond length distortion and bond angle distortion parameters respectively.

$$\Delta = \frac{1}{6}\sum_{i=1}^{6}\left(\frac{d_i - \bar{d}}{\bar{d}}\right)^2$$

$$\Sigma = \frac{1}{11}\sum_{i=1}^{12}|\varphi_i - 90|$$

In Pb(MMBA)$_2$, two extra closest S atoms are treated as ligands in the octahedron.

**Table S2.** Dephasing time of coherent optical phonon in some representative materials.

| Material | Main phonon energy (meV) | Dephasing time (ps) | Temperature (K) | Reference |
|---|---|---|---|---|
| 1L-MoS$_2$ | 50 | 1.7 | RT | 1 |
| 1L-WSe$_2$ | 30.8 | 4.5 | RT | 2 |
| 2H-MoTe$_2$ | 21 | 6.7 | 10 | 3 |
| LaAlO$_3$ | 30 | 14 | 6 | 4 |
| KMnF$_3$ | 40 | 70 | 6 | 4 |
| ZnO | 12.3 | 211 | 5 | 5 |
| Bi | 12 | 2.41 | RT | 6 |
| AgSePh | 12 | 1-2 | RT | 7 |
| AgSePh | 3.6-12.8 | 2.5-4.7 | 5 | 8 |
| (PEA)$_2$PbI$_4$ | 4.4 | 5.2 | 77 | 9 |
| Pb(MMBA)$_2$ | 2.9 / 3.3 | 7.4 / 17 | RT / 10 | This work |
| Pb(MSBT)$_2$ | 2.3 / 2.4 | 9.1 / 8.6 | RT / 10 | This work |
| Pb(MOBT)$_2$ | 0.73 / 1.0 | 15 / 75 | RT / 10 | This work |

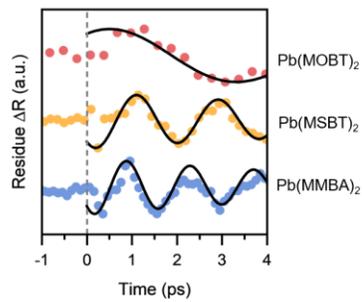

**Fig. S6**. Early time of the coherent phonon spectra at room temperature. All three LOC compounds exhibited cosine-type oscillation

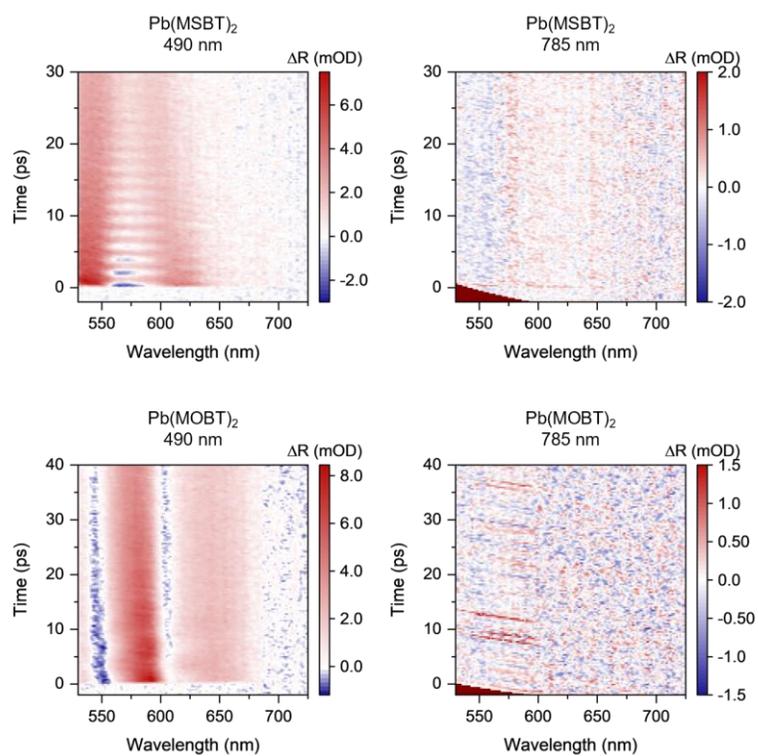

**Fig. S7**. TR spectra of Pb(MSBT)$_2$ and Pb(MOBT)$_2$ using resonant (490 nm, 810 uJ/cm$^2$) and non-resonant (785 nm 1600 uJ/cm$^2$) pump.

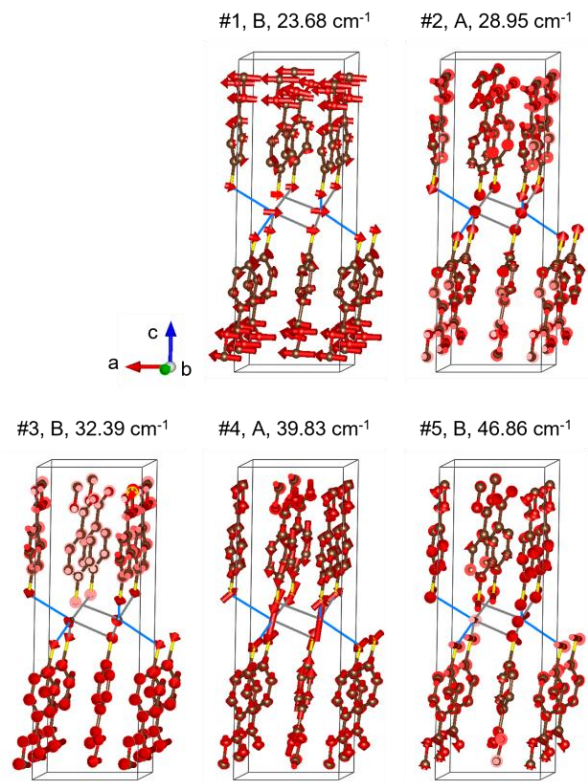

**Fig. S8**. Calculated low-energy phonon modes of Pb(MMBA)$_2$.

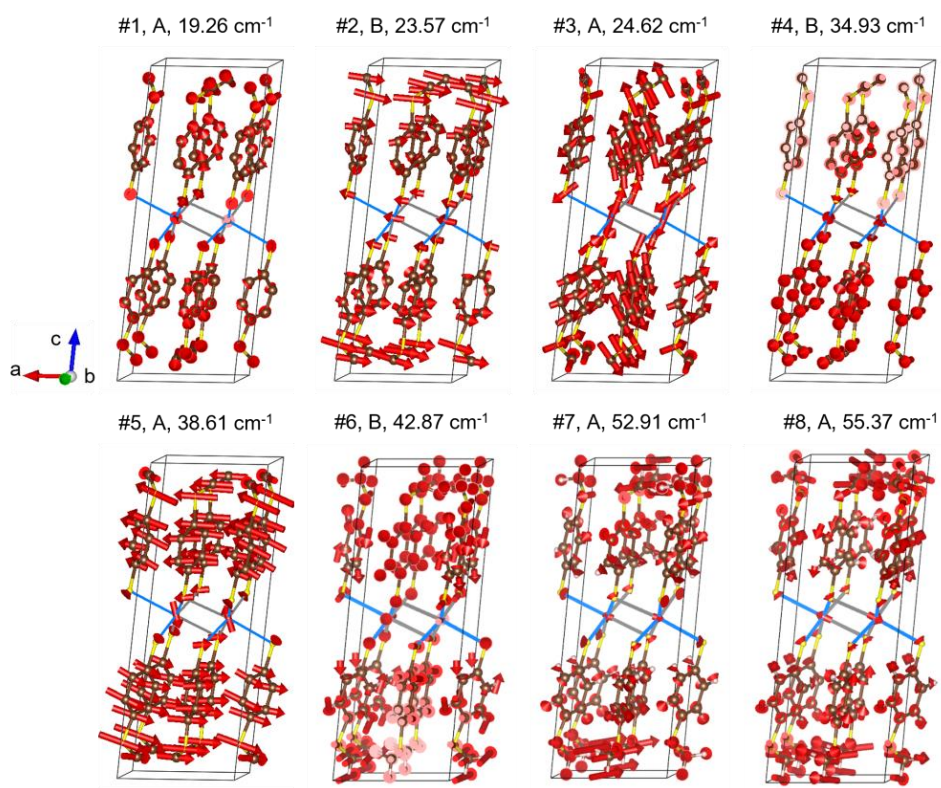

**Fig. S9**. Calculated low-energy phonon modes of Pb(MSBT)$_2$.

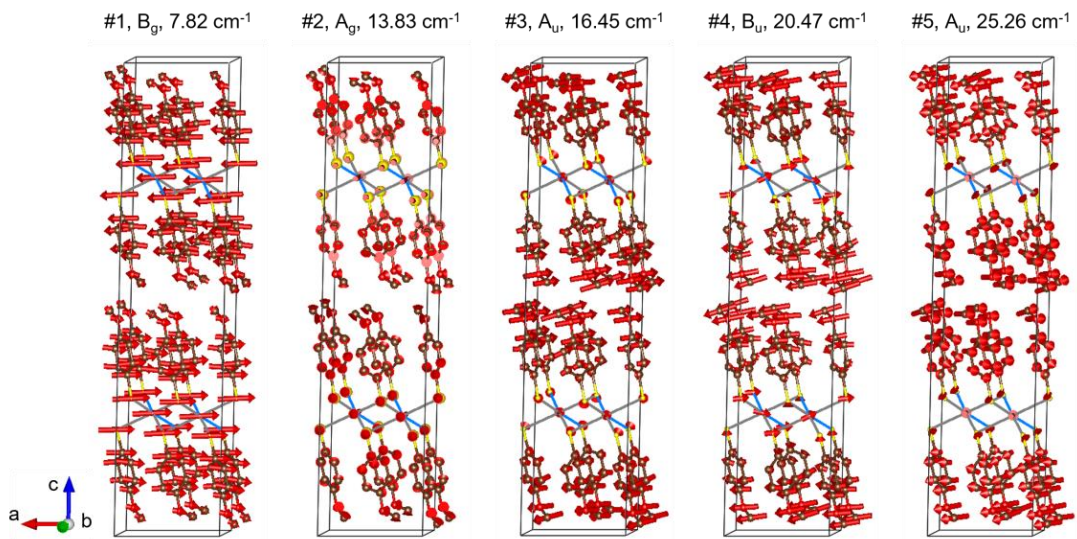

**Fig. S10**. Calculated low-energy phonon modes of Pb(MOBT)$_2$.

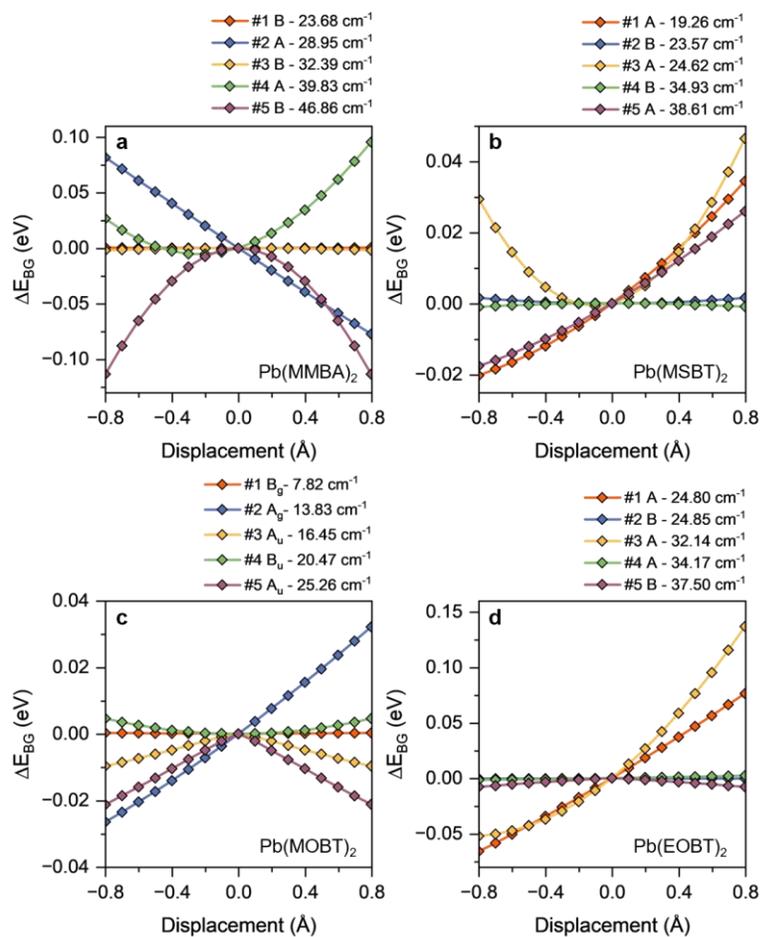

**Fig. S11**. Calculated band gap shift as a function of displacement along the eigenvector.

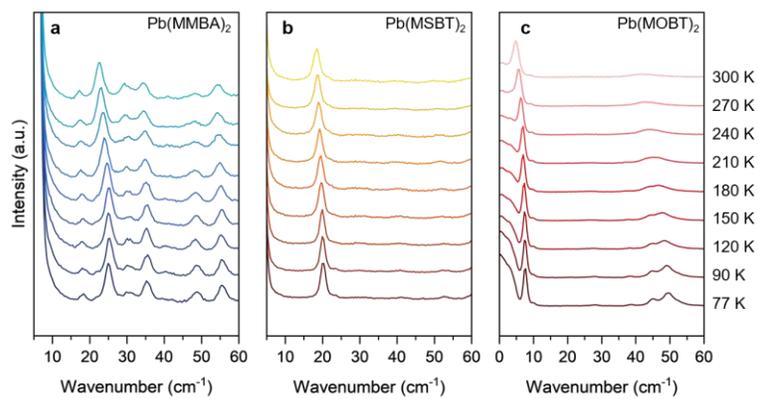

**Fig. S12**. Temperature-dependent Raman spectra of LOCs.

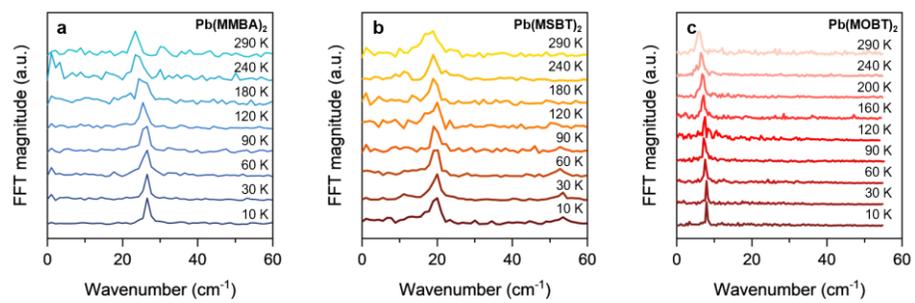

**Fig. S13**. Temperature-dependent FT spectra of TR signal.

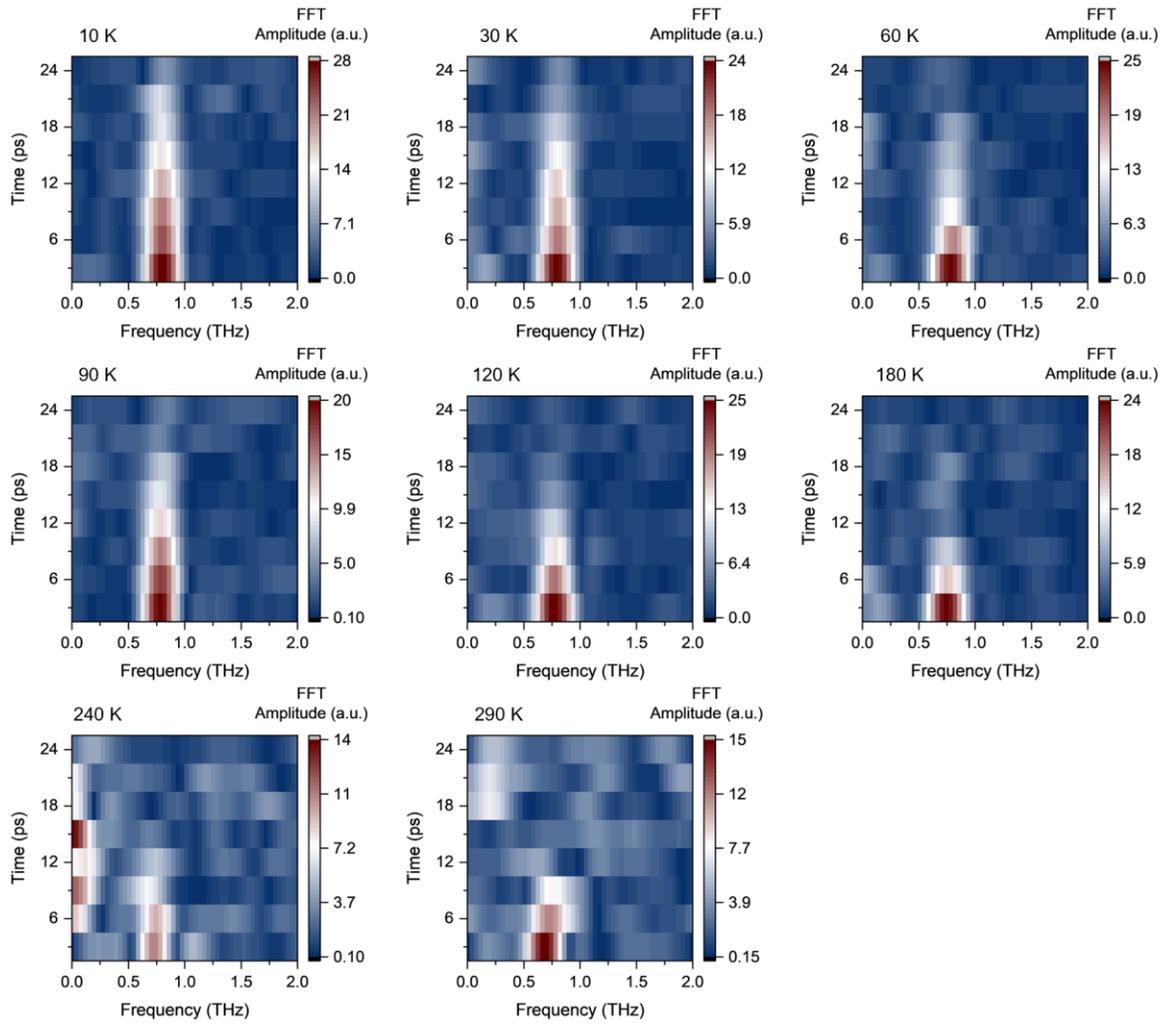

**Fig. S14.** Short time Fourier transform of the coherent phonon spectrum of Pb(MMBA)$_2$ with 0.1 ps resolution, 6 ps window and 50% overlap in adjacent window.

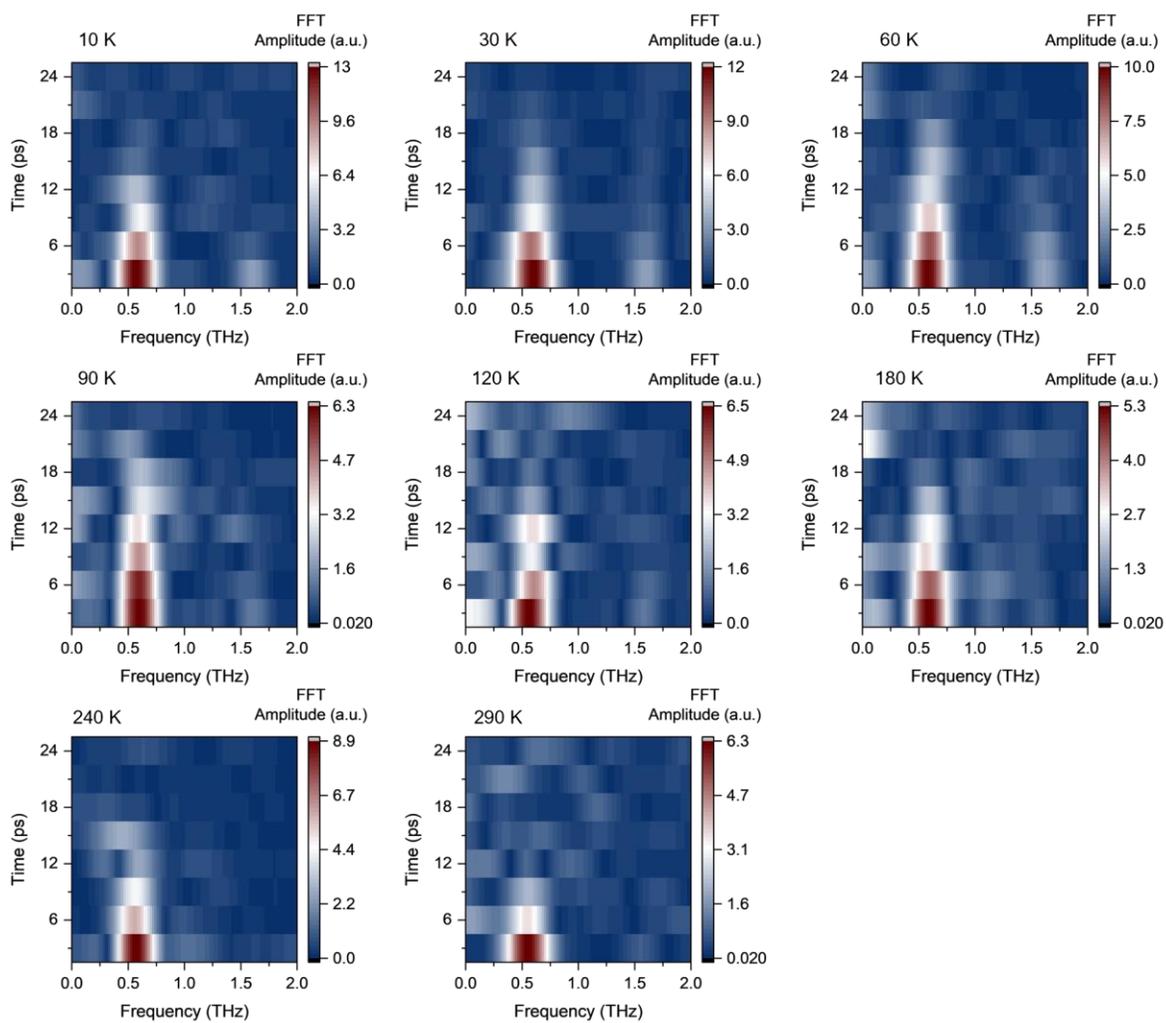

**Fig. S15**. Short time Fourier transform of the coherent phonon spectrum of Pb(MSBT)$_2$ with 0.2 ps resolution, 6 ps window and 50% overlap in adjacent window.

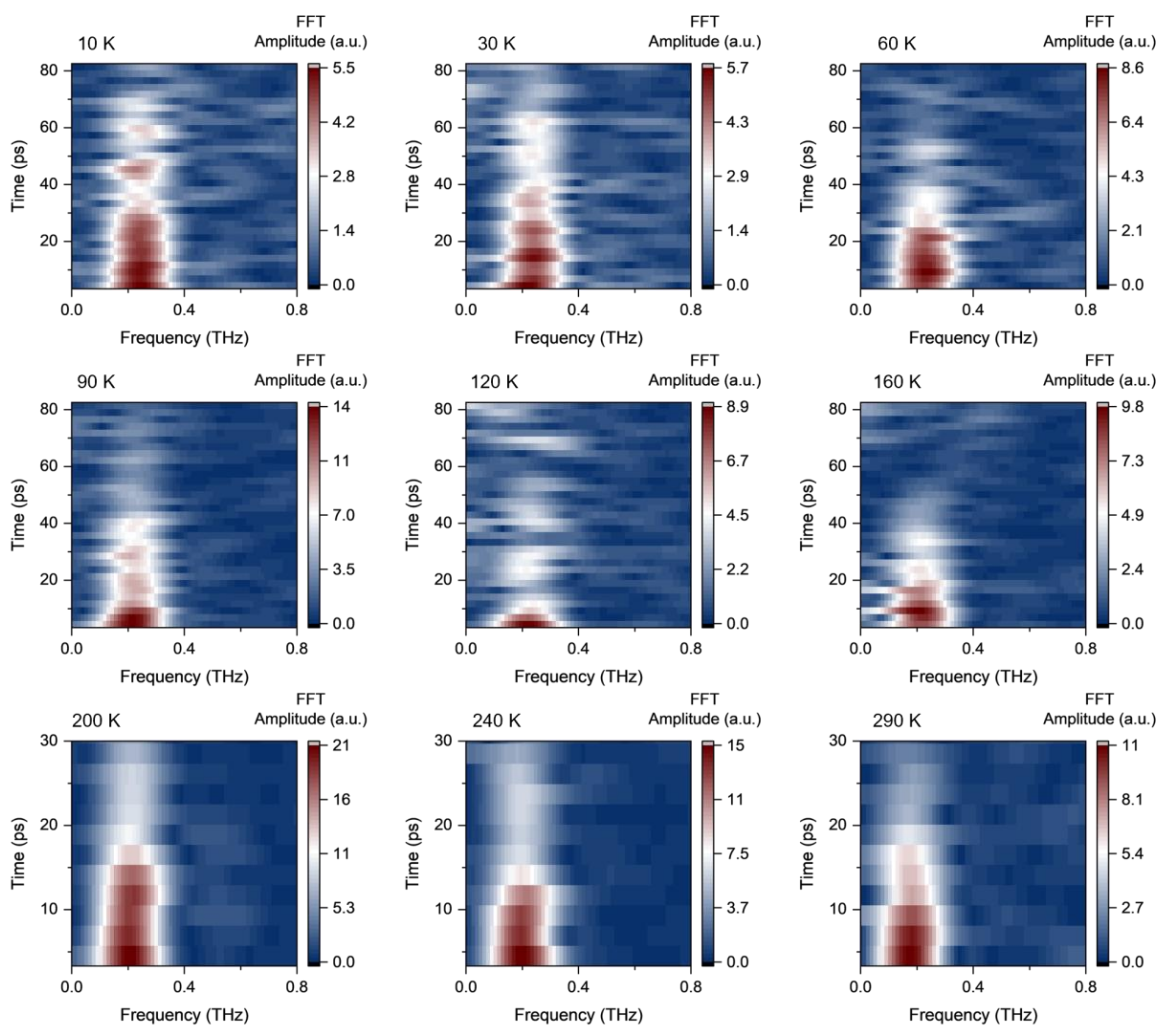

**Fig. S16**. Short time Fourier transform of the coherent phonon spectrum of Pb(MOBT)$_2$ with 0.3 ps resolution, 9 ps window and 75% overlap in adjacent window.

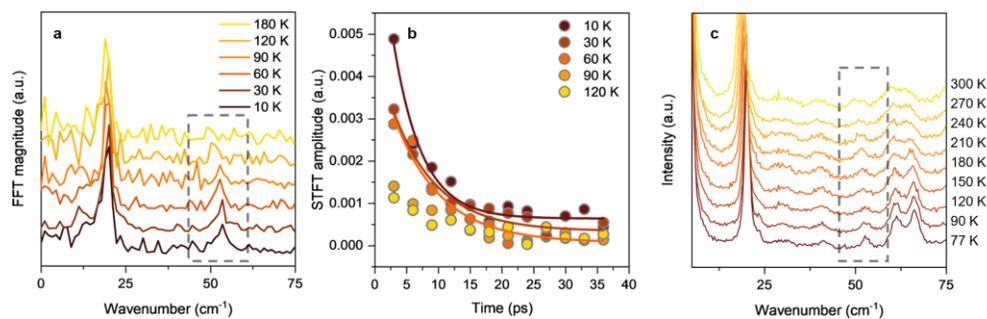

**Fig. S17**. **a.** Zoom-in view of temperature-dependent FT spectra of TR signal of Pb(MSBT)$_2$. **b.** Single exponent fitting of STFT amplitude at ~53cm$^{-1}$. **c.** Zoom-in view of temperature-dependent Raman spectra of Pb(MSBT)$_2$.

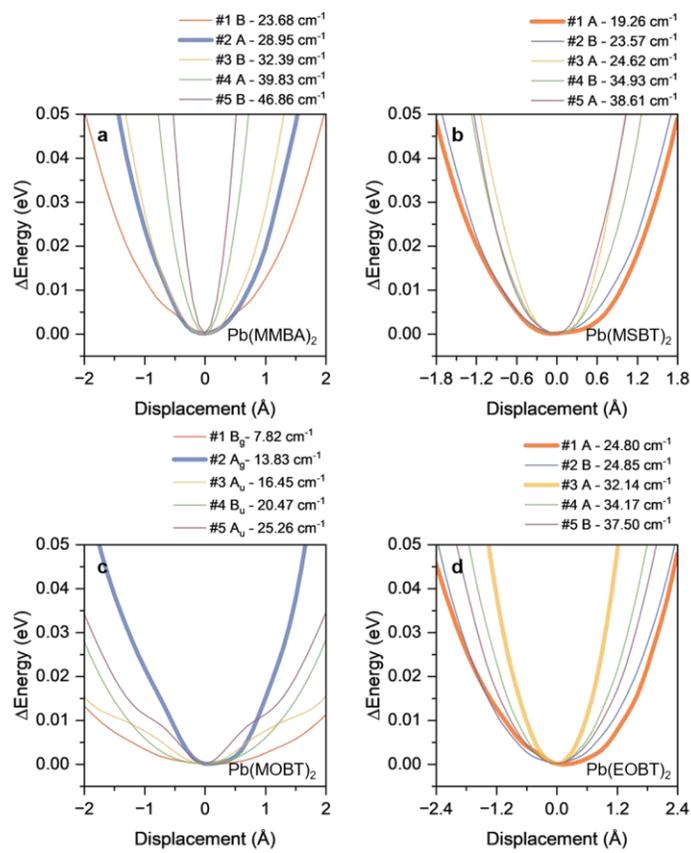

**Fig. S18**. Calculated vibrational energy potential of LOCs. The phonon modes probed in transient reflection spectroscopy are bold for clarity.

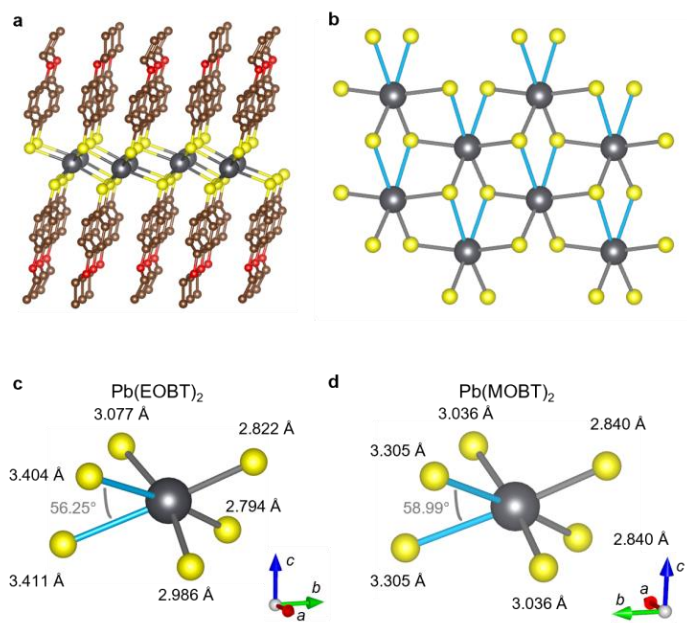

**Fig. S19**. **a**. Schematic illustrations of the crystal structure of Pb(EOBT)$_2$. **b**. In-plane Pb-S inorganic sublattice of Pb(EOBT)$_2$. **c, d**. Comparison of the PbS$_6$ octahedral in Pb(EOBT)$_2$ and Pb(MOBT)$_2$. Long Pb-S bonds (bond length > 3.1 Å and < 3.6 Å) are marked in blue in **b-d.**

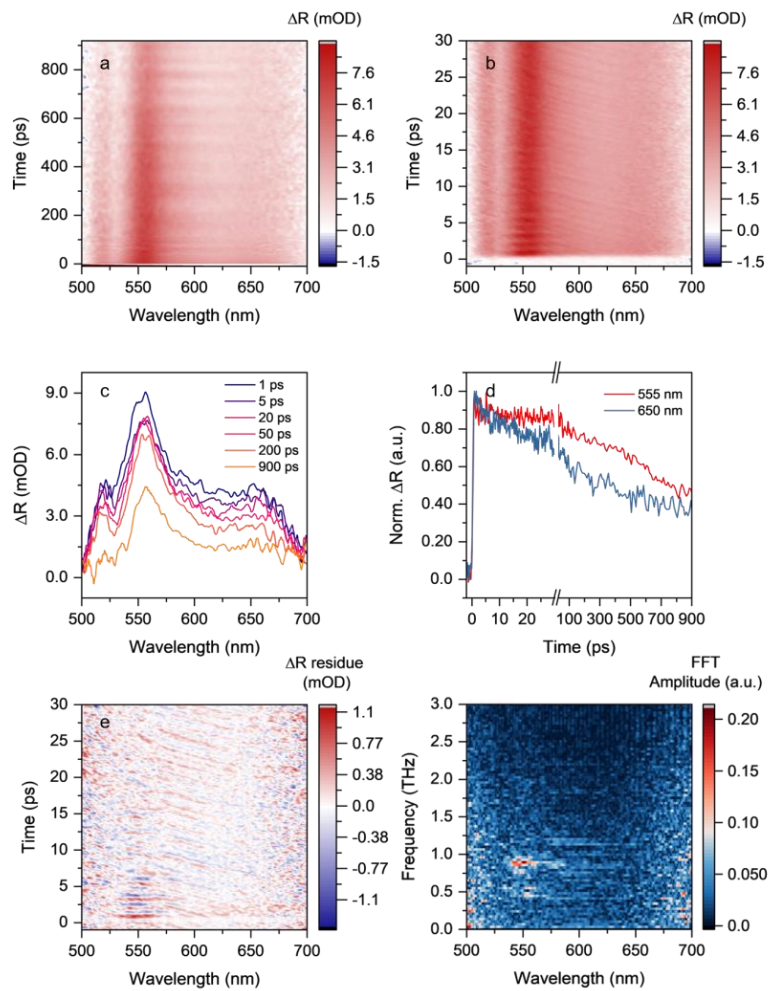

**Fig. S20. a, b.** Pseudo-color map of TR spectrum of Pb(EOBT)$_2$ at 10 K. **c**. TR spectrum of Pb(EOBT)$_2$ at different delay time. **d**. TR kinetic traces, probed at 555 nm and 650 nm. **e**. Pseudo-color mapping of residue TR signal. **f**. FFT mapping of the residue TR signal from 1 to 30 ps.

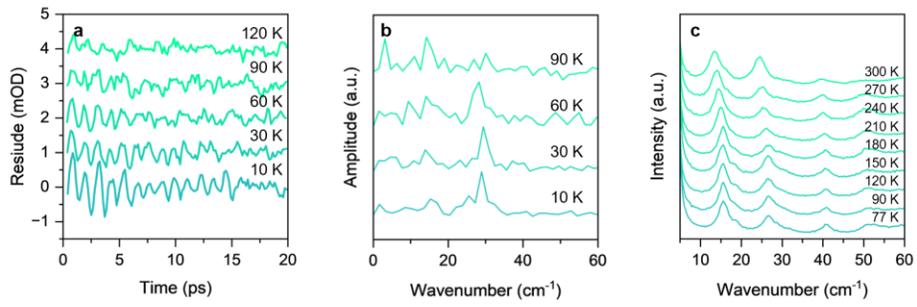

**Fig. S21**. **a.** Temperature-dependent residue of transient reflection spectrum of Pb(EOBT)$_2$. **b**. Fourier transform of coherent phonon spectrum between 0.5~15 ps. **c**. Temperature-dependent Raman spectra.

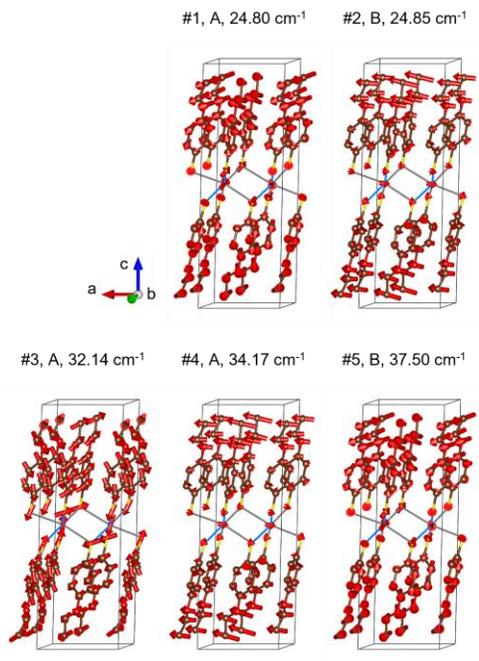

**Fig. S22**. Calculated low-energy phonon modes of Pb(EOBT)$_2$.

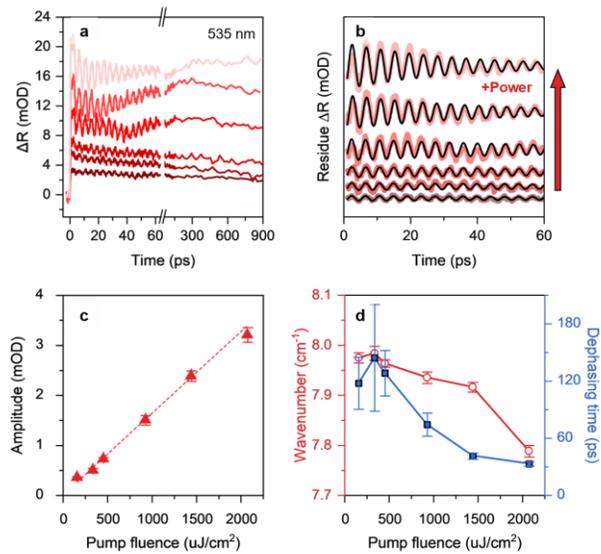

**Fig. S23**. **A.** Power dependent TR kinetics of Pb(MOBT)$_2$ at 10 K, probed at 535 nm. **b**. Power-dependent TR residue of Pb(MOBT)$_2$. Curves are offset for clarity. Exponential damped cosine fitting curves are marked in black lines. **c, d**. Summarized power-dependent amplitude, phonon energy and dephasing time.

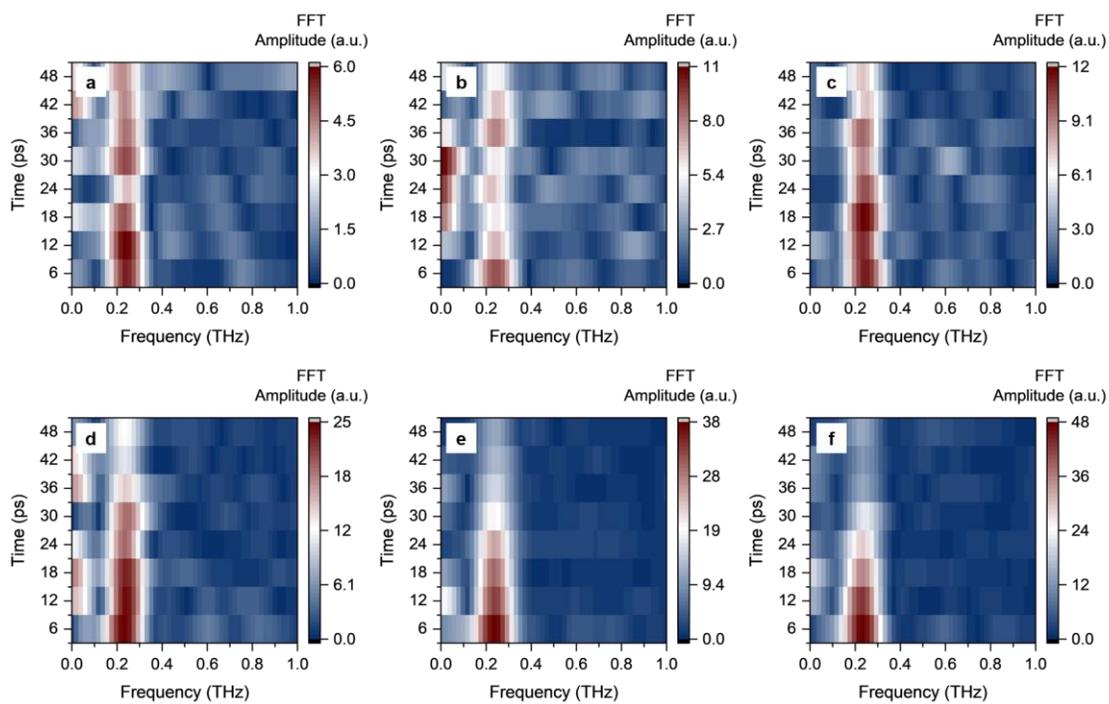

**Fig. S24.** Short time Fourier-transform of Pb(MOBT)$_2$ at 10 K at varying pump fluence of $1.6\times10^2$ uJ/cm$^2$, $3.4\times10^2$ uJ/cm$^2$, $4.5\times10^2$ uJ/cm$^2$, $9.3\times10^2$ uJ/cm$^2$, $1.4\times10^3$ uJ/cm$^2$, and $2.1\times10^3$ uJ/cm$^2$.